# Reconfigurable Intelligent Surface-Enabled Physical-Layer Network Coding for Higher Order M-QAM Signals

Ehsan Atefat Doost*, Firooz B. Saghezchi*, Pablo Fondo-Ferreiro†, Felipe Gil-Castiñeira†, Maria Papaioannou⸸, John Vardakas‡, Jinwara Surattanagul§, Jonathan Rodriguez§
*Instituto de Telecomunicações, Universidede de Aveiro, Aveiro, Portugal
†Information Technologies Group, atlanTTic Research Center, University of Vigo, Vigo, Spain
⸸Faculty of Engineering and Science, University of Greenwich, Chatham Maritime, UK
‡Iquadrat Informatica, S.L, Barcelona, Spain, and Dept. of Informatics, UOWM, Greece
§Faculty of Computing, Engineering and Science, University of South Wales, Pontypridd, UK
{e.atefatdoost, firooz}@av.it.pt, {pfondo, xil}@gti.uvigo.es, m.papaioannou@greenwich.ac.uk, jvardakas@iquadrat.com,
{jinwara.surattanagul, jonathan.rodriguez}@southwales.ac.uk

*Abstract*—Physical-Layer Network Coding (PNC) is an effective technique to improve the throughput and latency in wireless networks. However, there are two major challenges for PNC, especially when using higher order modulations: 1) phase synchronization and power control at the paired User Equipments (UEs); and 2) the ambiguity removal of the PNC mapping at the relay node. To address these challenges, in this paper, we apply power control at transmitting UEs and exploit Reconfigurable Intelligent Surfaces (RISs) to synchronize the phase of the transmitted signals and ensure that they arrive at the relay with the same power and phase rotation. Then, we employ modular addition for an unambiguous PNC mapping for M-ary Quadrature Amplitude Modulations (M-QAM). We evaluate the performance of the system in the framework of Orthogonal Frequency Division Multiplexing (OFDM)-PNC for different RIS sizes and modulation orders. Furthermore, we study the sensitivity of PNC systems for Channel Estimation Error (CEE). The results reveal that 1) PNC systems show quite higher sensitivity to CEE compared with RIS-assisted one-way relay channel systems; 2) when the CEE is low, RIS can considerably enhance the Signal-to-Noise Ratio (SNR) of the PNC system, e.g., for a Bit Error Rate (BER) of $10^{-3}$ (without channel coding), increasing the RIS size from one to 256 elements in 28 GHz band leads to 200% improvement in SNR.

## I. INTRODUCTION

Physical-layer Network Coding (PNC) [1], [2] is the enabling technique for bidirectional communications over Two-Way Relay Channels (TWRC). However, it requires phase synchronization and power control at the transmitting User Equipments (UEs) to ensure that the two signals arrive at the PNC relay node with equal power and phase rotation; otherwise, the constellation points of the two signals will intertwine and the superposed signal will end up having a short distance between its constellation points. This makes it very difficult (or even impossible) to reliably detect the signal at the relay, which leads to a high Bit Error Rate (BER) in the overall system performance.

However, this strict synchronization requirement by PNC may not be guaranteed in practical systems. Even if the two signals are transmitted simultaneously, they may arrive at the relay with different phase rotations, due to different fading channels between the UEs and the relay, e.g., due to their different distances from the relay and/or interactions with different scattering objects. The ambiguity of the PNC mapping at the relay is another challenge for PNC systems, especially when a non-binary signal constellation (e.g., M-QAM or M-PSK, with M ≥ 4) is adopted.

Previous research efforts addressing synchronization and power control issues in PNC systems [3-5] are mostly focused on a precoding technique where each of the two transmitting UEs multiplies their signal by the inverse of their respective channel response up to the relay node before its transmission. This ensures that the two signals arrive at the relay with an equal power and phase rotation, which maximizes the distance between the received constellation points at the relay node. In our previous work in [6], we extended these works to higher order M-QAM modulations in the framework of Orthogonal Frequency Division Multiplexing (OFDM)-PNC systems under both AWGN and fading channels. However, when the wireless channel experiences a deep fade, the channel inversion may end up in a prohibitively high transmit power, which may surpass the limit that the UE can accommodate.

Reconfigurable Intelligent Surface (RIS) is an emerging technology for Sixth Generation mobile networks (6G), which can improve the signal coverage and quality and cope with channel fading and radio blockages. It integrates many passive and tiny (sub-wavelength) reflecting elements on a linear array or a planar surface, whose phase shifts can be manipulated on demand and independently from one another through a Software-Defined Networking (SDN) controller to

This work was supported in part by i) FCT through the PhD scholarship with reference No. 2021.08263.BD; ii) Xunta de Galicia (Spain) under grant ED481B-2022-019; and iii) EXPLOR project funded by H2020-MSCA-RISE-2019 (grant agreement ID: 872897).

establish an *effective* Line-of-Sight (LoS) channel and circumvent radio blockages [7]. In fact, by tuning the phase shifts of these elements, the signals reradiated from these tiny elements can be added constructively to enhance the received signal power at the desired UE's location or can be combined destructively to mitigate the undesired signals at unintended UEs, e.g., to mitigate the multi-user interference and signal leakage to eavesdroppers.

RIS has been extensively explored in recent years and combined with other technologies, e.g., cell-free massive Multiple-Input Multiple-Output (massive-MIMO) [8], Simultaneous Wireless Information and Power Transfer (SWIPT) [9], Non-Orthogonal Multiple Access (NOMA) [10], Unmanned Aerial Vehicle (UAV) [11], and PNC [12], to enhance the performance of wireless networks. In particular, in PNC systems, RIS has been exploited in two different scenarios: 1) RIS as a PNC relay, where the active PNC relay is replaced with a passive RIS [13]; and 2) RIS-assisted active PNC relay, where the active PNC relay is maintained, but RISs are employed to enhance the channels between the UEs and the relay [14]. The former requires full duplex UEs and advanced signal processing to eliminate the self-loop interference. In contrast, the latter is compatible with half duplex radios since the PNC transmission and reception are performed over two separate time slots. Moreover, PNC systems can also be categorized from another perspective. The paired UEs may share a single RIS, or they may each employ a dedicated RIS. We refer to these scenarios as single-RIS and double-RIS PNC scenarios, respectively. It is worth noting that in a single-RIS scenario, as the RIS optimizes its phase shift vector for both UEs simultaneously, it inevitably leads to some performance degradation. In contrast, in a double-RIS scenario, there is more degrees of freedom and each RIS can be independently configured for one of the UEs.

In this paper, we adopt a double-RIS PNC scenario assisting an active PNC relay, as illustrated in Fig. 1. We offload the phase synchronization task from each of the two transmitting UEs to their corresponding RIS, hence relieving the UEs from performing any precoding operation for phase synchronization. The RIS indeed compensates the difference in phase rotation of the two signals, so as they both experience the same phase rotation over the channel, and hence they arrive at the active PNC relay node with aligned in-phase and quadrature components. In particular, the main contributions of this paper can be summarized as follows:

- Instead of using precoding for PNC phase synchronization in OFDM-PNC systems as proposed in [6], we adopt the idea of using RIS's phase shift capability for the synchronization of PNC symbols as proposed in [14] and extend it to OFDM-PNC framework [4] with power control mechanism to adjust the UEs' transmit powers ($P_A$ and $P_B$) to account for the different path losses seen by the two UEs.

- We implement a generalizable PNC mapping in the active PNC relay node that is compatible with Binary Phase-Shift Keying (BPSK) and readily scalable to higher-order M-QAM signals (e.g., 4-QAM, 16-QAM and 64-QAM).

- We assess the performance of the OFDM-PNC system for 4-QAM, and 16-QAM signals and for different RIS sizes (e.g., 64 and 256) and compare the results with a benchmark scenario where the RIS size is one (i.e., has only one reflecting element). Last but not least, we study the sensitivity of the OFDM-PNC system to the Channel Estimation Error (CEE).

The rest of this paper is organized as follows. Section II reviews the state of the art on RIS-assisted PNC systems. Section III presents our system model, including our adopted methods for the phase shift configuration of the RISs, power control at the paired UEs, and PNC mapping at the active PNC relay node, under both perfect and noisy channel estimations. Section IV describes our simulation setup and discusses the results. Finally, Section V concludes this paper and draws guidelines for the future work.

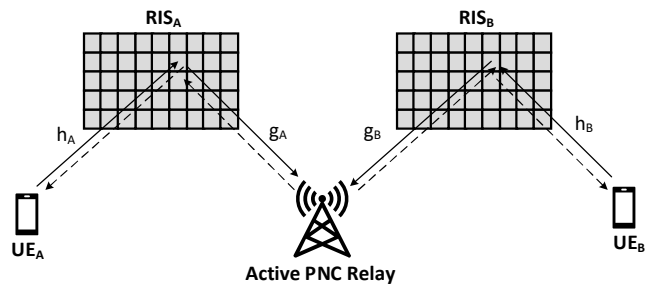

Fig. 1. Studied RIS-assisted PNC scenario with an active relay node and two RISs (The solid and dashed links represent the Multiple Access (MA) and the Broadcast (BC) phase of the PNC communications, respectively.)

II. RELATED WORK

Regarding the RIS-assisted active PNC relay scenario, few works exist in the literature. Tao et al., in [15], integrated RIS and Amplify-and-Forward (AF) relaying to maximize the end-to-end sum rate of a double-RIS PNC system, where they jointly optimize the phase shifts at both RISs. The authors of [16] analytically derived the BER in a double-RIS PNC system under a fading channel. In [14], the authors study both double-RIS and single-RIS PNC scenarios for a single carrier system, where they analyze the BER performance for BPSK modulation under a multipath fading channel. Generally, in a RIS-assisted PNC system, the phase matrix is a key factor which must be optimized to achieve a good performance, e.g., low BER [14] or high spectral efficiency [15].

Different from aforementioned works, in this paper, we combine RIS, OFDM-PNC, and PNC ambiguity resolving schemes to come up with a more robust solution for a multipath fading scenario and a scalable solution for an M-QAM signal with a higher constellation order (e.g., M=16 and 64) – that is yet backward compatible with the BPSK signal – to further push the spectral efficiency limits in PNC systems. For removing channel-induced Singular Fade States (SFSs) [17], we apply over-the-air synchronization using the RISs, which essentially transforms a multipath fading channel into an effective AWGN channel to better accommodate the non-binary QAM modulations (e.g., 4-QAM and 16-QAM). In addition, to remove the ambiguity of the PNC mapping, we use the modular addition techniques described in [6], which is scalable to high-order M-QAM signals and in the special case of M=2 (BPSK) it reduces to the conventional Exclusive OR (XOR) PNC mapping.

### III. SYSTEM MODEL

*A. PNC Signaling*

Fig. 1 illustrates our system model, where we assume that there are two single-antenna UEs (namely $UE_A$ and $UE_B$) that exchange data with each other through an active PNC relay node. We consider these communications take place in a millimeter-Wave (mmWave) band and in a dense radio environment (with frequent blockages), hence there is no direct link between the UEs and the active PNC relay node. Therefore, two RISs ($RIS_A$ and $RIS_B$) are employed to enhance the wireless channels between the UEs and the relay node. Each RIS has L reflecting elements, whose phase shifts can be configured with an infinite resolution. That is, they can take on any continuous value between 0 and $2\pi$.

For the PNC signaling, the two-way packet exchange is executed in two phases: 1) Multiple Access (MA) phase, and 2) Broadcast (BC) phase [1]. The process flow of these two phases is as follows.

**Multiple Access (MA) Phase:** In this phase, the paired UEs ($UE_A$ and $UE_B$) transmit their signals simultaneously to the relay node at the same subcarriers ($k=k_A=k_B$). All nodes are assumed to be half-duplex, i.e., the nodes cannot transmit and receive at the same time.

The received signal at the relay can be expressed as [14]:

$$y_R[n] = \sqrt{P_A}\ g_A^T\ \Theta_A\ h_A\ x_A[n] + \sqrt{P_B}\ g_B^T\ \Theta_B\ h_B\ x_B[n] + n_R[n]\ ;\ n=1,2,\ldots,N, \quad (1)$$

where $x_m$, $m \in \{A, B\}$, depicts the modulated symbols of $UE_m$, using the same M-QAM modulation, and $\Theta_m$, $m \in \{A,B\}$, is a diagonal matrix whose diagonal elements represent the amount of phase shifts applied by the reflecting elements of $RIS_m$ on its impinging signal, i.e.:

$$\Theta_m = \text{diag}\{\mu_{m,1}\ e^{j\theta_{m,1}}, \mu_{m,2}\ e^{j\theta_{m,2}}, \ldots, \mu_{m,L}\ e^{j\theta_{m,L}}\} \quad (2)$$

where $\mu_{m,i} \in [0, 1]$ and $\theta_{m,i} \in [0, 2\pi]$ represent, respectively, the amplitude attenuation and the phase shift introduced by the i-th reflecting element. In (2), $P_A$ and $P_B$ stand for the transmit powers of $UE_A$ and $UE_B$, respectively, which are adjusted according to the gain of their respected cascaded (UE-RIS-relay) channel. Furthermore, $n_R$ is an additive complex white Gaussian noise with zero mean and variance $\sigma_n^2$.

The channel vectors between $RIS_m$ and the relay is denoted as $g_m \in \mathbb{C}^{L \times 1}$, which is given as follows [14]:

$$g_m = \sqrt{pl}\ g_m^f,\ m \in \{A, B\}, \quad (3)$$

Where $pl$ is the channel path loss and $g_m^f$ is the multipath fading term. Similarly, the channel vector between $UE_m$ and $RIS_m$ is denoted as $h_m \in \mathbb{C}^{L \times 1}$, which is modelled as follows:

$$h_m = \sqrt{pl}\ h_m^f,\ m \in \{A, B\} \quad (4)$$

where $h_m^f$ represents the fading component. We use *simRIS* channel model proposed in [18] and implemented in [19].

We use RIS for synchronizing the two streams of symbols transmitted by $UE_A$ and $UE_B$ [14]. The phase shift coefficients of the two RISs ($\Theta_A$ and $\Theta_B$) are adjusted to compensate for 1) the different phase rotations of $x_A$ and $x_B$ reflected from different reflecting elements so that they all arrive at the PNC relay with a similar phase rotation; 2) the relative phase rotation of $x_A$ and $x_B$ due to the independent fading that they experience on their channel to the relay. The latter in particular can result in a so-called SFSs that cause distance shortening in the signal constellation received at the relay [6]. Furthermore, to confine the transmitted powers of the UEs to a maximum limit and ensure that the two signals arrive at the relay with an equal power, we propose a power control mechanism at the UEs to adjust their transmit power ($P_A$ and $P_B$) to compensate for their different channel attenuations.

However, CEE is unavoidable in wireless systems due to, e.g., channel aging and measurement noise. Therefore, we assume that the channel gain of the i-th reflecting element is given by $\hat{g}_{m,i} = g_{m,i} + e_{CE}^{\hat{g}}$ ($\hat{h}_{m,i} = h_{m,i} + e_{CE}^{\hat{h}}$), where $g_{m,i}$ ($h_{m,i}$) is the true value of the channel coefficient and $e_{CE}^{\hat{g}}$ ($e_{CE}^{\hat{h}}$) is the CEE. The CEE terms are two independent complex Gaussian random variables with same statistics, i.e., $e_{CE}^{\hat{g}} \sim \mathcal{CN}(0, \sigma_{CEE}^2)$ and $e_{CE}^{\hat{h}} \sim \mathcal{CN}(0, \sigma_{CEE}^2)$.

In order to compensate for the overall phase rotation of the cascaded RIS channel, the phase shift of the i-th reflecting element of $RIS_m$ should be configured as follows:

$$\hat{\theta}_{m,i} = -(\theta_{\hat{h}_{m,i}} + \theta_{\hat{g}_{m,i}}) \quad (5)$$

where $\theta_{\hat{h}_{m,i}} = \angle \hat{h}_{m,i}$ and $\theta_{\hat{g}_{m,i}} = \angle \hat{g}_{m,i}$ denote the angles of the estimated channel gains $\hat{h}_{m,i}$ and $\hat{g}_{m,i}$, respectively.

Assuming that the RIS controller has the Channel State Information (CSI) knowledge of the cascaded RIS channels, it can calculate the optimum phase shift coefficients of the reflecting elements using (5). Therefore, (1) can be rewritten as follows:

$$y_R[n] = \sqrt{P_A}\left(\sum_{i=1}^{L} g_{A,i}\mu_{A,i}\ e^{j\hat{\theta}_{A,i}}\ h_{A,i}\right) x_A[n] + \sqrt{P_B} \left(\sum_{i=1}^{L} g_{B,i}\mu_{B,i}\ e^{j\hat{\theta}_{B,i}}\ h_{B,i}\right) x_B[n] + n_R[n];\ n=1,2,\ldots,N. \quad (6)$$

Denoting the two summation terms of this equation by $\alpha_A$ and $\alpha_B$ ($\alpha_A = \sum_{i=1}^{L} g_{A,i}\mu_{A,i}\ e^{j\hat{\theta}_{A,i}}\ h_{A,i}$ and $\alpha_B = \sum_{i=1}^{L} g_{B,i}\mu_{B,i}\ e^{j\hat{\theta}_{B,i}}\ h_{B,i}$) yields:

$$y_R[n] = \sqrt{P_A}\ \alpha_A\ x_A[n] + \sqrt{P_B}\ \alpha_B\ x_B[n] + n_R[n],\ n=1,2,\ldots,N. \quad (7)$$

In the ideal case of no CEE ($\sigma_{CEE}^2 = 0$), one can easily confirm that $\alpha_A$ and $\alpha_B$ become real coefficients as follows:

$$\alpha_A = \sum_{i=1}^{N} |g_{A,i}|\ \mu_{A,i}\ |h_{A,i}| \quad (8)$$

$$\alpha_B = \sum_{i=1}^{N} |g_{B,i}|\ \mu_{B,i}\ |h_{B,i}|, \quad (9)$$

We assume that each round of PNC signaling (MA+BC) is performed within one channel coherence time. Hence, the channel reciprocity holds for the forward and reverse channels between the UEs and relay, i.e. $\alpha_A=\alpha_{R\text{-RIS-}A}=\alpha_{A\text{-RIS-}R}$ and $\alpha_B=\alpha_{R\text{-RIS-}B}=\alpha_{B\text{-RIS-}R}$ for all subcarriers $k=1,\ldots, N)$.

At the relay node, after removing the Cyclic Prefix (CP) and performing the N-point Fast Fourier Transform (FFT) on each OFDM symbol, the received signal in the frequency (constellation) domain becomes:

$$Y_R[k]=\sqrt{P_A}\,\alpha_A X_A[k]+\sqrt{P_B}\,\alpha_B X_B[k]+N_R[k]; \; k=1,2,\ldots,N. \quad (10)$$

We can see from this equation that in order to ensure an equal signal power arrival at the PNC relay node for the two superposed signals, the UEs' transmit powers must satisfy:

$$\sqrt{P_A}\,\alpha_A=\sqrt{P_B}\,\alpha_B. \quad (11)$$

Without loss of generality, we assume that $|\alpha_A| \geq |\alpha_B|$. Since $X_A$ and $X_B$ carry the same order of M-QAM modulation and $E(|X_A[k]|^2)=E(|X_B[k]|^2)=1$, to censure power control, the following inequality must hold.

$$P_A \leq P_B \leq P_{max} \quad (12)$$

To respect the maximum transmit power limit of the UEs, we propose the following power control mechanism. Assuming $P_B=P_{max}$, both magnitude and phase equalities should hold to satisfy (11). That is,

$$\angle \alpha_A = \angle \alpha_B \quad (13)$$

$$P_A\,|\alpha_A|^2=P_B\,|\alpha_B|^2 \quad (14)$$

Introducing a new variable $\gamma = |\alpha_B|/|\alpha_A|$, we will have:

$$P_A=\gamma^2\,P_{max} \quad (15)$$

That is, using this proposed power control mechanism, the UE with the weakest channel gain, transmits with its maximum power ($P_{max}$) and the other UE, which is experiencing a better channel, transmits with a fraction ($\gamma^2 \leq 1$) of $P_{max}$.

Using (13) and (14), (10) becomes:

$$Y_R[k]=\gamma^2 P_{max}\left(e^{j\angle \alpha_A}X_A[k]+e^{j\angle \alpha_B}X_B[k]\right)+N_R[k] \quad (16)$$

**Broadcast (BC) Phase:** In the second phase (time slot) of the PNC signaling (i.e., the BC phase), the relay node performs PNC mapping on each arrived symbol (in every subcarrier), converting $Y_R[k]$ to $X_R[k]$. Since the power control mechanism ensures that the amplitude of the signals of each paired UEs arriving at the relay are equal, we can adopt the PNC mapping scheme presented in [20]. After this PNC mapping, the relay broadcasts the network-coded data ($X_R[k]$) toward both UEs on the same subcarriers used in the MA phase. The received signal at each UE is then becomes.

$$Y_A[k]=\sqrt{P_R}\,\alpha_A\,X_R[k]+N_A[k] \quad (17)$$

$$Y_B[k]=\sqrt{P_R}\,\alpha_B\,X_R[k]+N_B[k] \quad (18)$$

Where, $P_R$ denotes the transmit power of the relay node.

At the UEs, each of the paired UEs detects $\hat{Z}=\hat{f}(X_A[k], X_B[k])=(X_A[k]+X_B[k])\bmod \sqrt{M}$, where the modulus ($\sqrt{M}$) is equal to the square root of the M-QAM signal's constellation size, assuming a square-shape QAM constellation. Here, the modular $\sqrt{M}$ addition is to remove the ambiguity of the PNC mapping and detect the signal of the peer. Further discussion on PNC ambiguity removal is out of the scope of this paper. Interested readers can refer to [6] for further discussions. Finally, after the detection of $\hat{Z}$, $UE_A$ ($UE_B$) calculates the mod $\sqrt{M}$ addition of its own transmitted symbol $X_A[k]$ ($X_B[k]$) and $\hat{Z}$ to obtain the modulation symbol transmitted by its peer.

## IV. PERFORMANCE EVALUATION RESULTS

Since the end-to-end BER of PNC systems is dominated by the BER of the uplink phase, due to the synchronization and PNC mapping errors, we assess only the BER contribution of the uplink phase in our simulations. We use design parameters according to IEEE 802.11p standard [21] as summarized in Table I. For the channel model, we consider a realistic model with pathloss effects where the channel coefficients are modelled as in [18], [19], considering physical and environmental effects.

TABLE I. SIMULATION PARAMETERS

| Parameters | Value | |
|---|---|---|
| Bandwidth | 10 MHz for OFDM symbol | |
| Number of antennas | At relay | 1 |
| | Per UE | 1 |
| Number of RIS elements (L) | Per RIS | 1, 4, 16, 64, 256 |
| Node coordinates (x,y,z) in meter | $UE_A$ | (0,2,1.5) |
| | $UE_B$ | (0,30,1.5) |
| | $RIS_1$ | (0,8,2.5) |
| | $RIS_2$ | (0,22,2.5) |
| | Relay | (0,14,2) |
| Number of UEs | 2 | |
| Carrier frequency ($f_c$) | 28 GHz | |
| Channel model for $g_j$ and $h_j$ | Rayleigh | |
| Subcarrier spacing ($\Delta f=B/N$) | 156.25 KHz | |
| $T_{FFT}=4*T_{CP}=T_{symbol}-T_{CP}$ | 64 | 6.4 µs |
| Cyclic prefix ($T_{CP}=1/(4*\Delta f)$) | 16 | 1.6 µs |
| Symbol length ($T_{symbol}=T_{FFT}+T_{CP}=5*T_{CP}$) | 80 | 8 µs |
| Number of data Subcarriers | 48 | |
| Number of pilot subcarriers | 4 | |
| Total Number of subcarriers | 64 | |
| Modulation | 4-QAM, 16-QAM | |
| Baseband (FFT) sampling rate | 10M samples/s (10 MHz) | |
| Data Rate | 24 Mbps (16-QAM) | |

We evaluate the RIS-enabled scenario for different RIS sizes and M-QAM modulation orders, and compare them against a benchmark scenario when the RIS size is 1 (i.e., the RIS has only one reflecting element).

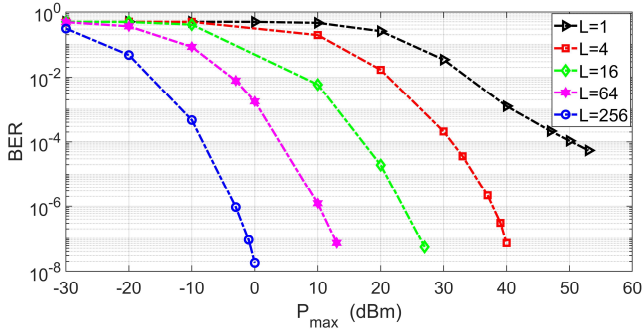

Fig. 2. BER of the 4-QAM for different RIS sizes and transmit powers.

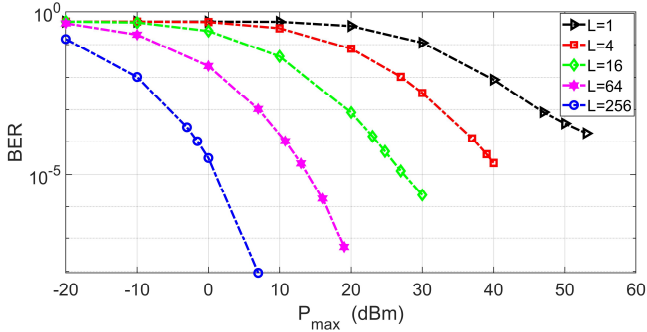

Fig. 3. BER of 16-QAM for different RIS sizes and transmit powers.

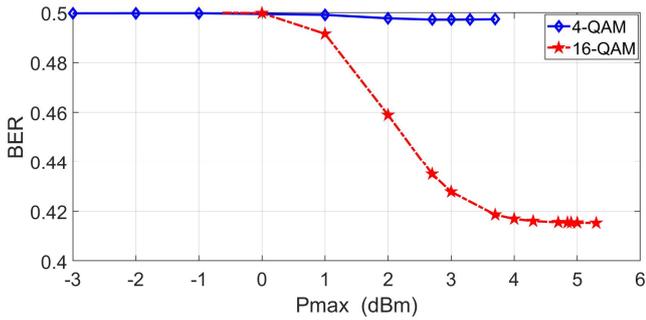

Fig. 4. BER of 4-QAM and 16-QAM for a RIS size of L=16 using random phase shifts.

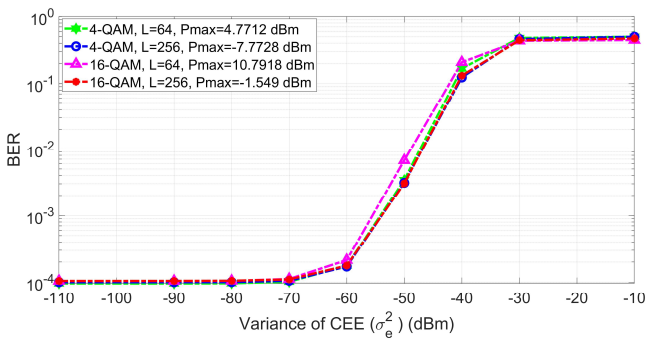

Fig. 5. Sensitivity of the BER to CEE for the considered PNC system for 4-QAM and 16-QAM under two different RIS sizes (L=64, 256).

We chose this benchmark since we can generate its channel model in a similar fashion as for other RIS sizes to have a fair comparison. It is worth mentioning that we can even view this benchmark as a scenario where the RIS is replaced with a random scatterer and the phase synchronization is performed at UEs, rather than at the RISs.

Figs. 2 and 3 depict the BER performance for 4-QAM and 16-QAM, respectively, when the RIS size increases from 1 to 256, the proposed power control mechanism is used, and the phase shifts of reflecting elements are configured to their optimal values, given by (5). Overall, for both 4-QAM and 16-QAM, using a larger RIS requires less transmit power, for a given BER. In particular, we observe from Fig. 3 that for 16-QAM to achieve a BER of $10^{-4}$, the system requires around -1.5, 11, 24, 37 and 55 dBm transmit power ($P_{max}$) for the RIS sizes of 256, 64, 16, 4, and 1, respectively.

In contrast to Figs. 2 and 3, where the phase shifts of RIS elements are optimal, we conducted another experiment (with L=16) where we set the phase shifts of RIS elements randomly. Fig. 4 illustrates the results for both 4-QAM and 16-QAM. We observe from this figure that PNC is very sensitive to phase errors and does not work with random phase shifts since as seen in the figure, the BER is always about 0.5.

Evaluating the impact of imperfect CSI on BER, Fig. 5 shows the variations of the BER for 4-QAM and 16-QAM signals and two different RIS sizes (L=64 and 256) when the CEE increases from a -110 dBm to -10 dBm. Note that the transmit powers are different for the four curves, as shown in the legend. We intentionally set these different transmit power values so as all four curves start from BER=$10^{-4}$ when CEE=-110 dBm. We observed that our proposed PNC system attains the same BER-CEE variance performance regardless of the RIS size and modulation order. This issue highlights the merit of the proposed synchronous OFDM-PNC system in terms of its adaptability to different modulation orders or different RIS sizes. However, the sensitivity of the signal detection algorithm in the OFDM-PNC system to the variance of the CEE is still a legitimate concern. As Fig. 5 shows the channel estimation variance should be confined within -60 dBm for all the scenarios of different RIS sizes and different modulation orders, as shown in the results.

Using near real-time channel estimation and tracking the channels behaviors, the synchronization and power control mechanism used in this work can perfectly remove channel artifacts and achieve a robust operation. Otherwise, there would be an error floor for high CEEs.

V. CONCLUSION AND FUTURE WORK

In this paper, we addressed the phase offset compensation in OFDM-PNC systems by employing RIS and power control for M-QAM modulations. The results suggest that for non-binary M-QAM modulations (M>=4), over-the-air phase compensation at RIS as well as power control at the paired UEs are crucial to avoid the ambiguities stemming from the phase distortions of the MA channels. This enables the relay to successfully detect superposed M-QAM symbols. We also observe that, unlike conventional RIS-assisted one-way relay channel systems that can operate even with random phase shifts, PNC is quite sensitive to phase errors and is easily disrupted with random phase shift configurations. Furthermore, when we have accurate and timely CSI, RIS can considerably enhance the SNR in PNC systems. For example, for BER=$10^{-3}$ and without channel coding, increasing the RIS size from one to 256 in 28GHz band triples the SNR (i.e., a 200% improvement).

For future work, this work can be extended in two different directions: 1) employing massive-MIMO for RIS-assisted multi-user OFDM-PNC communications; and 2) employing UAVs to carry the RISs and/or the active PNC relay node to dynamically optimize their location in order to further enhance the system performance.